\documentclass[amsmath,a4,11pt]{article}

\date{\empty}

\begin{document}

\title{{\bf Superadiabatic-type magnetic amplification in
conventional cosmology}}

\author{Christos G. Tsagas${}^1$\thanks{e-mail address:
c.tsagas@damtp.cam.ac.uk} and Alejandra Kandus${}^2$\thanks{email
address: kandus@uesc.br}\\ {\small ${}^1$DAMTP, Centre for
Mathematical Sciences, University of Cambridge}\\ {\small
Wilberforce Road, Cambridge CB3 0WA, UK}\\ {\small
${}^2$LATO-DCET, Universidade Estadual de Santa Cruz, Rodovia
Ilh\'eus-Itabuna km 16 s/n}\\ {\small  Salobrinho CEP-05508-900,
Ilh\'eus - BA, Brazil}}

\maketitle

\begin{abstract}
We consider the evolution of cosmological magnetic fields in FRW
models and outline a geometrical mechanism for their
superadiabatic amplification on large scales. The mechanism
operates within standard electromagnetic theory and applies to FRW
universes with open spatial sections. We discuss the general
relativistic nature of the effect and show how it modifies the
adiabatic magnetic evolution. Assuming a universe that is only
marginally open today, we estimate the main features of the
superadiabatically amplified residual field.
\end{abstract}

\section{Introduction}
Magnetic fields appear everywhere in the universe. From Earth and
the nearby stars, all the way to the remote galaxy clusters and
high-redshift protogalaxies, the existence of magnetic fields has
been repeatedly verified~\cite{K}. Despite this widespread
presence, however, the origin of cosmic magnetism remains a
mystery and it is still the subject of debate. Over the years,
numerous mechanisms of magnetogenesis have appeared in the
literature (see~\cite{En} for recent reviews). Broadly speaking,
one can classify these scenarios into those arguing for a late
(post-recombination) magnetic generation and those advocating a
primordial origin for the field~\cite{H}. Prior to recombination
we have superstring and inflation based models, mechanisms
operating during the electroweak and the quark-hadron phase
transitions, eddies in the pre-recombination plasma and effects at
electron-proton recombination. These early time scenarios are
mainly global amplification mechanisms of primeval magnetic seeds.
In the post-recombination era there exist local astrophysical
processes that generate the magnetic seeds and operate
simultaneously with the amplifying process. For example, weak
magnetic fields produced via the Biermann battery~\cite{Biermann}
can be amplified to galactic size fields during the protogalactic
collapse~\cite{KCOR}. Alternatively, stronger magnetic seeds can
be injected into the intragalactic medium by stellar winds and
supernova explosions (e.g.~see~\cite{Rees}). In addition, there is
still relatively little knowledge on the reionization of our
universe and the related magnetohydrodynamics.

An attractive aspect of early magnetogenesis is that it makes the
ubiquity of large-scale magnetic fields in the universe,
particularly those observed in high-redshift protogalaxies, easier
to explain. Inflation seems the most plausible candidate for
producing the primordial fields, as it naturally leads to
large-scale phenomena from subhorizon microphysics. The main
obstacle in this scenario is that any early magnetic field that
survives an epoch of inflation is so drastically diluted that it
can never seed the galactic dynamo for the ordered, large-scale
field.\footnote{In the galaxy there is also a random magnetic
field, which is stirred up by turbulent motions and saturates on
scales of $\sim50-100$~pc. This field has magnitude slightly
larger than that of the ordered magnetic field but its growth time
is only $3\times10^7$~yrs (i.e.~one tenth of the typical growth
timescale associated with the ordered field dynamo)} The reason is
cosmological magnetic flux conservation, namely the fact that the
strength of the large-scale fields drops as $a^{-2}$ ($a$ is the
scale factor of the universe). The root of the problem is traced
down to the conformal invariance of electromagnetism and to the
conformal flatness of the Friedmann-Robertson-Walker (FRW) models.
Together, these two are thought to guarantee that $B\propto
a^{-2}$ always and irrespective of plasma effects. When the FRW
background has nonzero spatially curvature, however, we will show
this is not necessarily the case.

The most common way of modifying the `adiabatic' $B\propto a^{-2}$
law is by breaking away from standard electromagnetic theory.
There are more than one ways of doing that, which explains the
large number of relevant scenarios in the literature. Perhaps the
first detailed discussion of the issue was the one given
in~\cite{TW}. Among other suggestions, the authors introduced a
coupling between the Maxwell field and the curvature of the space
in their Lagrangian. As a result, both the conformal invariance
and the gauge invariance of Maxwell's equations were lost.
However, when applied to a spatially flat FRW universe, the
aforementioned interaction led to an extra magneto-curvature term
in the magnetic wave equation The immediate consequence was that
superhorizon-sized magnetic fields, evolving in a poorly
conducting inflationary universe, decayed slower than the standard
$a^{-2}$-law. This meant an effective superadiabatic amplification
of the field on these scales, a concept that was originally
introduced in gravitational wave studies~\cite{G}. In other words,
magnetic fields on large enough scales could go through an epoch
of inflation and still remain strong enough to sustain the
galactic dynamo.

Following~\cite{TW}, several mechanisms producing magnetic fields
during inflation and reheating have appeared in the
literature~\cite{Ga}. Most of these scenarios break the conformal
invariance of the Maxwell field by introducing extra couplings
with the spacetime curvature, or non-conformally invariant sources
to Maxwell's equations. In this paper we consider a conventional
interaction between the electromagnetic and the gravitational
field, which so far has been sparsely studied in cosmology. This
is the natural general relativistic coupling between
electromagnetism and spacetime geometry that emerges from the
vector nature of the Maxwell field and from the geometrical
approach of Einstein's theory. The best known effect of the
aforementioned interaction, which emerges from the Ricci
identities, is probably the `scattering' of electromagnetic
radiation by the gravitational field~\cite{DWB}. In what follows
we will show that, under certain circumstances, the same coupling
can also lead to the superadiabatic amplification of cosmological
magnetic fields without violating or modifying standard
electromagnetism. Our mechanism operates primarily on magnetic
fields coherent on the largest subcurvature scales of a spatially
open FRW universe, which asymptotically approaches flatness as it
undergoes a period of inflationary expansion. The result is that
these fields decay as $a^{-1}$, a rate considerably slower than
the adiabatic $a^{_2}$-law. Therefore, primordial magnetic fields
that survive an epoch of inflation could be considerably stronger
than previously anticipated due to curvature effects alone. In
practise this means that magnetic fields which are coherent on
very large scales could have appreciable strengths. Assuming that
$1-\Omega\sim10^{-2}$ today, in particular, we find a residual
field of $10^{-35}$~G spanning a comoving length of
$\sim10^4$~Mpc. This is much stronger than any other large-scale
field obtained by conventional methods. Moreover, in a universe
currently dominated by a dark energy component, a seed field of
$10^{-35}$~G lies within the broad galactic dynamo
requirements~\cite{DLT}.

The attractive aspects of the mechanism presented below are its
simplicity and the fact that it operates within standard
electromagnetic theory. If the universe is marginally open today,
this scenario could provide a viable method for a superadiabatic
type of early magnetic amplification  and lead to fields with
astrophysically interesting strengths on very large scales. Even
if the universe is not open, however, our mechanism still offers a
simple general relativistic counter-example to the widespread
perception that the superadiabatic amplification of magnetic
fields in FRW cosmologies is  not possible within conventional
electromagnetism. In either case, we believe that this study will
further facilitate our theoretical understanding of the subject,
while it may also prove a valuable step in the ongoing quest for
an answer to the origin of cosmic magnetism.

\section{Magnetic fields in curved FRW universes}
We begin by reminding the reader that, with the exception of
completely random radiation or a fully tangled magnetic field,
electromagnetic fields are not compatible with the highly
symmetric FRW spacetimes. The isotropy of the latter means that
these models cannot naturally accommodate inherently anisotropic
sources like the Maxwell field. The implication is that, strictly
speaking, even weak cosmological electromagnetic fields should be
studied in perturbed Friedmann models. Here we will also show that
the standard magnetic evolution of $B\propto a^{-2}$ does not
always hold in Friedmann models with nontrivial spatial curvature.
All these mean that studying cosmological magnetic fields in flat
Minkowski space is a good approximation only when the fields are
weak and only on small scales in models with nontrivial spatial
curvature. In the latter case the approximation becomes
progressively less accurate as one moves to larger scales and the
3-curvature effects start kicking in. Technically speaking, this
means that certain linear couplings between the field and the
geometry of the 3-space, which vanish only when the background is
identically flat, are bypassed. It is the purpose of this paper to
examine the implications of these magneto-geometrical couplings
for the evolution of cosmological magnetic fields.

Our analysis uses the covariant approach to general relativity,
which introduces a family of timelike fundamental observers moving
with 4-velocity $u_a$ (i.e.~$u_au^a=-1$). We assume that relative
to $u_a$ the cosmic medium has a perfect fluid form with a
barotropic equation of state, and that the fundamental observers
experience an electromagnetic field with components $E_a$ and
$B_a$. Both $E_a$ and the pseudovector $B_a$ live on the
observers' local rest space (i.e.~$E_au^a=0=B_au^a$). In the
absence of vorticity, the projection tensor
$h_{ab}=g_{ab}+u_au_a$, where $g_{ab}$ is the spacetime metric, is
also the metric of the spatial hypersurfaces. The electromagnetic
field obeys the standard Maxwell's formulae, consisting of two
propagation equations\footnote{Angled brackets denote spatially
projected vectors and the projected, symmetric and trace-free part
of spacelike second-rank tensors (e.g.~$\dot{B}_{\langle
a\rangle}=h_a{}^b\dot{B}_b$). Also, round brackets indicate
symmetrisation and square ones antisymmetrisation.}
\begin{eqnarray}
\dot{B}_{\langle a\rangle}&=&-{\textstyle{2\over3}}\Theta B_a+
\left(\sigma_{ab}+\varepsilon_{abc}\omega^c\right)B^b-
\varepsilon_{abc}\dot{u}^bE^c- {\rm curl}E_a\,,  \label{eq:M1}\\
\dot{E}_{\langle a\rangle}&=&-{\textstyle{2\over3}}\Theta E_a+
\left(\sigma_{ab}+\varepsilon_{abc}\omega^c\right)E^b+
\varepsilon_{abc}\dot{u}^bB^c+ {\rm curl}B_a- {\cal J}_a\,,
\label{eq:M2}
\end{eqnarray}
with $\Theta$ representing the volume expansion, $\sigma_{ab}$ the
shear, $\omega_a$ the vorticity, $\dot{u}_a$ the 4-acceleration
and ${\cal J}_a=J_{\langle a\rangle}$ the projected
4-current~\cite{E}. The above are supplemented by the constraints
\begin{eqnarray}
{\rm D}^aB_a&=&2\omega^aE_a\,,  \label{eq:M3}\\
{\rm D}^aE_a&=&\rho_{\rm e}- 2\omega^aB_a\,,  \label{eq:M4}
\end{eqnarray}
where $\rho_{\rm e}$ is the charge density. Note that overdots
indicate proper time derivatives and ${\rm D}_a=h_a{}^b\nabla_b$
is the covariant derivative operator on the observer's local
3-space. Also, ${\rm curl}v_a=\epsilon_{abc}{\rm D}^bv^c$ for any
orthogonally projected vector $v_a$ (i.e.~with $v_au^a=0$) and
$\epsilon_{abc}$ is the projected permutation tensor. By
differentiating Eq.~(\ref{eq:M1}) with respect to time and then
using (\ref{eq:M2}) to eliminate $E_a$, one arrives at the
covariant wave equation of $B_a$ in a general spacetime~\cite{T1}.
Linearised about a FRW background the latter reads
\begin{equation}
\ddot{B}_a- {\rm D}^2B_a= -5H\dot{B}_a- 4H^2B_a+
{\textstyle{1\over3}}\rho(1+3w)B_a- {\cal R}_{ab}B^b+ {\rm
curl}{\cal J}_a\,,  \label{eq:Bwave1}
\end{equation}
where $H=\Theta/3=\dot{a}/a$ is the background Hubble parameter,
$\rho$ is the energy density of the matter and $w=p/\rho$, where
$p=p(\rho)$ is the barotropic pressure. When linearising the full
equations we assume that the magnetic field vanishes in the
unperturbed FRW background. This guarantees the gauge-invariance
of the analysis and frees our results from any gauge related
ambiguities (see~\cite{T1} for further discussion and technical
details). Note the second last term in the right-hand side of the
above, where ${\cal R}_{ab}=(2k/a^2)h_{ab}$ is the zero-order
spatial Ricci tensor and $k=0,\pm1$ is the associated curvature
index. This term is manifestly linear and vanishes only when the
background model is spatially flat. Cosmological magnetic field
studies in flat spaces will clearly bypass such
magneto-geometrical terms. The latter result from the general
relativistic coupling between the electromagnetic and the
gravitational field and are an unavoidable consequence of the
geometrical nature of Einstein's theory. Technically speaking this
magneto-geometrical interaction is manifested in the 3-Ricci
identity. In the absence of rotation, the latter reads $2{\rm
D}_{[c}{\rm D}_{b]}B_a={\cal R}_{dabc}B^d$, where ${\cal
R}_{abcd}$ represents the spatial Riemann tensor and ${\cal
R}_{ab}={\cal R}^c{}_{acb}$~\cite{T1,TB2}. In what follows we will
consider the implications of the magneto-curvature term in the
right-hand side of Eq.~(\ref{eq:Bwave1}) for the evolution of
cosmological magnetic fields.

The effect of the current term in the right-hand side of
(\ref{eq:Bwave1}) depends crucially on the conducting properties
of the medium in which the magnetic field evolves. If Ohm's law
holds, then the electrical conductivity is the quantity that
describes these properties. In general there are additional terms
in what is known as the generalized Ohm's law. For example, when
building a magneto-hydrodynamical model of three separate fluids,
namely electrons, protons and neutrals, the interaction of the
first two gives rise to the Hall effect and the last two lead to
ambipolar diffusion (e.g.~see~\cite{gen-ohm}). However, only the
resistive term is responsible for the dissipation of the magnetic
energy into heat, while the other effects do not cause
dissipation. The Hall term might have influenced the evolution of
the field during the radiation era~\cite{Tajima} and the ambipolar
diffusion is effective in the intragalactic
medium~\cite{Sano,Zweibel}. During inflation the universe is
normally treated as a very poor conductor. Thus, Ohm's law
guarantees that all spatial currents vanish, despite the presence
of nonzero electric fields (e.g.~see~\cite{T1}). Given that we are
primarily interested in the evolution of a large-scale primordial
magnetic field during a early period of inflation, we will from
now on ignore the current contribution to
Eq.~(\ref{eq:Bwave1}).\footnote{On sufficiently large scales the
current term in Eq.~(\ref{eq:Bwave1}) is negligible even during
the standard Big-Bang evolution. Indeed, by definition ${\rm
curl}{\cal J}_a=\epsilon_{abc}{\rm D}^b{\cal J}^c=
\epsilon_{abc}\partial^b{\cal J}^c$, given the symmetry of the of
Christoffel symbols. Moreover, $\partial_a{\cal J}_b\sim{\cal
J}/L$, where ${\cal J}^2={\cal J}_a{\cal J}^a$ and $L$ is the
scale in question. Clearly, as we move to progressively larger
wavelengths $\partial_a{\cal J}_b\rightarrow0$.}

After inflation, the reheating process reinstates the high
electrical conductivity of the cosmic medium. Of course, the
resistivity of the plasma is not identically zero. Nevertheless,
the amount of magnetic dissipation on large scales is negligible.
We can estimate the decay time of a magnetic field coherent over a
scale $L$ (with $L$ smaller than the horizon scale) as $t_{\rm
d}\sim L^2/\lambda$, where $\lambda$ is the magnetic diffusivity
(e.g.~see~\cite{edcm}). Assuming Spitzer conductivity at an epoch
when the temperature of the universe is $T\sim10^4$ K, we have
$\lambda\sim10^{7}~{\rm cm}^2/{\rm sec}$ and obtain $t_{\rm d}\sim
10^{26}~\rm{yrs}$ for fields coherent on approximately 100~pc.
This timescale is many orders of magnitude larger than the current
age of the universe. Therefore, the magnetic flux on
astrophysically interesting scales is effectively frozen into the
cosmic plasma.

We adopt the standard decomposition $B_a=B_{({\rm n})}Q_a^{({\rm
n})}$, where $Q_a^{({\rm n})}$ is the ${\rm n}$-th vector
harmonic, ${\rm D}_aB_{({\rm n})}=0=\dot{Q}_a^{({\rm n})}$, ${\rm
D}^aQ_a^{({\rm n})}=0$ and ${\rm D}^2Q_a^{({\rm n})}=-({\rm
n}^2/a^2)Q_a^{({\rm n})}$. Then, substituting the background
expression of ${\cal R}_{ab}$ into Eq.~(\ref{eq:Bwave1}) we obtain
\begin{equation}
\ddot{B}_{({\rm n})}+ 5H\dot{B}_{({\rm n})}+ 4H^2B_{({\rm n})}-
{\textstyle{1\over3}}\rho(1+3w)B_{({\rm n})}+ {2k\over
a^2}B_{({\rm n})}+ {{\rm n}^2\over a^2}B_{({\rm n})}=0\,,
\label{eq:Bwave2}
\end{equation}
for the evolution of the ${\rm n}$-th magnetic mode. The Laplacian
eigenvalues take continuous values, with ${\rm n}^2\geq0$, when
$k=0,-1$ and discrete ones, with ${\rm n}^2\geq3$, for $k=+1$. In
this notation supercurvature modes in spatially open models have
$0\leq{\rm n}^2<1$, which guarantees that the physical wavelength
of the perturbation is larger than the curvature scale
(i.e.~$\lambda_{\rm n}=a/{\rm n}>a$). On the other hand, modes
with ${\rm n}^2>1$ span lengths smaller than the curvature scale
and will be therefore termed subcurvature. Note that the
supercurvature modes are always larger than the Hubble length and
consequently never in causal contact. On the other hand,
perturbations on subcurvature scales can be causally connected
(see~\cite{LW} for further discussion). For our purposes, the
causality of magnetic modes with ${\rm n}^2>1$ is crucial.
Finally, we remind the reader that ${\rm n}^2=0$ denotes the
so-called homogeneous mode.

To proceed further we recall that the zero-order Raychaudhuri
equation does not explicitly depend on the background curvature
and takes the form $\ddot{a}/a=-\rho(1+w)/6$. On using this
expression and introducing $\eta$, the conformal time variable
with $\dot{\eta}=1/a>0$, Eq.~(\ref{eq:Bwave2}) becomes
\begin{equation}
B_{({\rm n})}''+ 4\left({a'\over a}\right)B_{({\rm n})}'+
2\left({a'\over a}\right)^2B_{({\rm n})}+ 2\left({a''\over
a}\right)B_{({\rm n})}+ 2kB_{({\rm n})}+ {\rm n}^2B_{({\rm
n})}=0\,,  \label{eq:Bwave3}
\end{equation}
where primes indicate differentiation with respect to $\eta$.
Finally, employing the `magnetic flux' variable ${\cal B}_{({\rm
n})}=a^2B_{({\rm n})}$ the above reduces to
\begin{equation}
{\cal B}_{({\rm n})}''+ {\rm n}^2{\cal B}_{({\rm n})}= -2k{\cal
B}_{({\rm n})}\,.  \label{eq:Bwave4}
\end{equation}
This wave equation shares a very close resemblance with the one
obtained in~\cite{TW} (see Eq.~(2.15) there). The similarity is in
the presence of a curvature related source term in both
expressions. The difference is that here the magneto-curvature
term is a natural and unavoidable consequence of the vector nature
of the magnetic field and of the geometrical approach of general
relativity. No new physics has been introduced and standard
electromagnetism still holds.

\section{The superadiabatically amplified magnetic field}
For a spatially flat background, the magneto-curvature term in
Eq.~(\ref{eq:Bwave4}) vanishes and one recovers the standard
wave-like evolution of the field, with an amplitude decreasing
according to the familiar $a^{-2}$-law. The adiabatic depletion
rate is also preserved when the background is spatially closed,
despite the presence of a non-zero magneto-curvature term in
(\ref{eq:Bwave4}). Indeed, for $k=+1$ the latter exhibits an
oscillatory solution of the form~\cite{T1}
\begin{equation}
B_{({\rm n})}=\frac{1}{a^2}\left\{{\cal C}_1\cos\left[\sqrt{{\rm
n}^2+2}\,\eta\right]+ {\cal C}_2\sin\left[\sqrt{{\rm
n}^2+2}\,\eta\right]\right\}\,,  \label{+1B}
\end{equation}
for the ${\rm n}$-th magnetic mode (with ${\cal C}_1$, ${\cal
C}_2$ constants). Apart from modifying the oscillation frequency,
the magneto-curvature term in Eq.~(\ref{eq:Bwave4}) has no
significant effect on the evolution of the field when $k=+1$. Note
that in this case the oscillatory behaviour of the magnetic field
is ensured on all scales by the closed geometry (i.e.~by the
compactness) of the space.

When dealing with the hyperbolic geometry of a spatially open FRW
model, however, the oscillatory behaviour of $B_{({\rm n})}$ is
not always guaranteed. Indeed, for $k=-1$ Eq.~(\ref{eq:Bwave4})
reads
\begin{equation}
{\cal B}_{({\rm n})}''+\left({\rm n}^2-2\right){\cal B}_{(n)}=0\,,
\label{eq:-1Bwave}
\end{equation}
which clearly does not accept an oscillatory solution on
sufficiently long wavelengths (i.e.~for ${\rm n}^2<2$). These
wavelengths extend from large subcurvature scales, with $1\leq{\rm
n}^2<2$, to supercurvature lengths with $0\leq{\rm n}^2<1$. Let us
consider the largest subcurvature scales first, since on these
wavelengths the associated magnetic modes can be causally
connected. It is convenient to introduce the parameter ${\rm
k}^2=2-{\rm n}^2$, so that the range $0<{\rm k}^2\leq1$
corresponds to the largest subcurvature scales. Then,
Eq.~(\ref{eq:-1Bwave}) assumes the form
\begin{equation}
{\cal B}_{({\rm k})}''- {\rm k}^2{\cal B}_{({\rm k})}=0\,,
\label{eq:-1Bwave1}
\end{equation}
yielding the following solution for large-scale magnetic fields
\begin{equation}
B_{({\rm k})}={1\over a^2}\left[{\cal C}_1\cosh(|{\rm k}|\eta)+
{\cal C}_2\sinh(|{\rm k}|\eta)\right]\,.  \label{eq:-1B}
\end{equation}
On these scales the standard $B\propto a^{-2}$ law is not a priori
guaranteed. Indeed, consider a FRW universe with open spatial
sections. Then, the Friedmann and the Raychaudhuri equations
combine to provide the expression
\begin{equation}
aH=\coth\left[{\textstyle{1\over2}}(1+3w)\eta+C\right]\,.
\label{eq:aH}
\end{equation}
where $C$ depends on the normalisation. The above governs the
expansion dynamics during the various epochs in the lifetime of
this universe, provided that the barotropic index $w$ remains
constant throughout each period. For our purposes the key period
is that of an inflationary expansion with $p/\rho=w=-1$. The
reason is that then the conductivity of the cosmic medium is
effectively zero and the magnetic evolution is monitored by
Eqs.~(\ref{eq:Bwave2})-(\ref{eq:Bwave4}). Also, the most dramatic
suppression of the field occurs during inflation and therefore any
change in the magnetic depletion rate during that period could
prove crucial. Note that inflation does not change the geometry of
the 3-space, but simply makes it look flatter by pushing the
curvature scale well beyond the observer's horizon. Setting $C=0$,
which means that $\eta<0$, reduces (\ref{eq:aH}) to
$aH=-\coth\eta$. The latter integrates to give
\begin{equation}
a=\frac{A_0e^{\eta}}{1-e^{2\eta}}\,, \label{eq:etarad}
\end{equation}
with $A_0=a_0(1-e^{2\eta_0})/e^{\eta_0}$ a positive constant
(see~\cite{T1} for details).\footnote{The adopted normalisation
scheme, where $C=0$ and $\eta<0$, has allowed us to streamline the
key equations considerably without loss of generality. Within
these conventions, $a\rightarrow0$ for $\eta\rightarrow-\infty$
and $a\rightarrow+\infty$ as $\eta\rightarrow0^-$.} Substituting
this result into the right-hand side of Eq.~(\ref{eq:-1B}) we can
express the evolution of the magnetic field in terms of the
cosmological scale factor. For simplicity consider the case of
$|{\rm k}|\rightarrow1^-$, which corresponds to the largest
subcurvature scales with ${\rm n}^2\rightarrow1^+$. Then, from
(\ref{eq:-1B}) and (\ref{eq:etarad}) we arrive at
\begin{equation}
B={\cal C}_3\left(1-e^{2\eta}\right)a^{-1}+ {\cal C}_4
e^{-\eta}a^{-2}\,,  \label{eq:-1Bf}
\end{equation}
where ${\cal C}_3$ and ${\cal C}_4$ are constants. Therefore, on
the largest subcurvature scales, the dominant magnetic mode never
depletes faster than $a^{-1}$. This decay rate is considerably
slower than the typical $a^{-2}$-law and holds throughout the
inflationary era. Note that the magnetic depletion switches to the
adiabatic $a^{-2}$ rate at the $\eta\rightarrow0^-$ limit
only.\footnote{According to Eq.~(\ref{eq:-1Bwave}), the curvature
effects modify the magnetic evolution on large scales with ${\rm
n}^2<2$. Expression (\ref{eq:-1Bf}) shows that as $|{\rm
k}|\rightarrow1^-$, which corresponds to ${\rm n}^2\rightarrow1^+$
and the largest subcurvature scales, the magnetic field decays as
$a^{-1}$. When ${\rm n}^2\rightarrow2^-$, on the other hand, we
have $|{\rm k}|\rightarrow0^+$ and $B\propto a^{-2}$. In
particular, expressions (\ref{eq:-1B}), (\ref{eq:etarad}) combine
to provide the general solution
\begin{equation}
B_{({\rm k})}={\cal C}_3\left(1-e^{2\eta}\right)^{|{\rm k}|}
a^{|{\rm k}|-2}+ {\cal C}_4\left(1-e^{2\eta}\right)^{-|{\rm k}|}
a^{-|{\rm k}|-2}\,,  \label{eq:-1Bg}
\end{equation}
with $|{\rm k}|\leq1$. Clearly, when $|{\rm k}|$ takes its values
in the open interval (0,1) the decay rate of the dominant magnetic
mode varies between $a^{-2}$ and $a^{-1}$, which is always slower
than the adiabatic $a^{-2}$-law.} Result (\ref{eq:-1Bf})
immediately implies that, beyond a certain scale, the cosmological
magnetic flux increases with time instead of being preserved.
Hence, in spatially open almost-FRW universes, large-scale
magnetic fields that survive inflation are significantly stronger
than anticipated because of curvature effects alone.

\section{The residual magnetic field}
In the previous sections we have studied the evolution of
large-scale primordial magnetic fields, emphasising on their
behaviour during the inflationary regime of a spatially open FRW
cosmology. So far we have provided a qualitative analysis that
identified a superadiabatic-type amplification for magnetic fields
spanning the largest subcurvature scales of the universe. Next we
will attempt to estimate the key properties of these
superadiabatically amplified fields, namely their strength and
coherence length.

Following~\cite{TW}, the energy density stored in the ${\rm n}$-th
magnetic mode as it crosses outside the horizon is $\rho_{\rm
B}=(M/m_{\rm Pl})^4\rho$, where $\rho\simeq M^4$ is the total
energy density of the universe and $m_{\rm Pl}$ is the Planck
mass. Then, assuming that $B^2\propto a^{-4}$, the energy density
in the mode at the end of the inflationary regime is given
by~\cite{TW}
\begin{equation}
\rho_{\rm B}=\frac{B^2}{8\pi}\sim
10^{-104}\tilde{\lambda}^{-4}_{\rm Mpc}\rho_{\gamma}\,.
\label{eq:r1}
\end{equation}
Here $\rho_{\gamma}$ is the radiation energy density and
$\tilde{\lambda}$ is the comoving scale of the field. The latter
is measured in Mpcs and it is normalised so that $\tilde{\lambda}$
coincides with the physical scale today. Note that the magnetic
mode crossed outside the horizon $N=N(\tilde{\lambda})$ e-folds
before the end of inflation (see~\cite{TW} for details). The
underlying assumption leading to the above result is that any
given mode is excited quantum mechanically while inside the
horizon and `freezes in' as a classical perturbation once it
crosses through the Hubble radius. The dramatic weakness of the
residual field demonstrated in Eq.~(\ref{eq:r1}), reflects the
drastic suppression of the magnetic energy density relative to the
vacuum energy, which remains constant throughout the inflationary
regime. After inflation $\rho_{\gamma}$ also decays as $a^{-4}$
and the ratio $r=\rho_{\rm B}/\rho_{\gamma}$ does not change.

If the dynamo amplification of large-scale fields is efficient,
the strength of the required magnetic seed, as measured at the
time of completed galaxy formation, ranges from $\sim10^{-19}$~G
down to $\sim10^{-23}$~G. In addition, the coherence length of the
initial field should be at least as large as the size of the
largest turbulent eddy, namely no less than $\sim100$~pc. The
aforementioned magnetic strengths, which correspond to
$r\sim10^{-27}$ and $r\sim10^{-35}$ respectively, have been
obtained in a spatially flat universe with zero cosmological
constant. However, if the universe is open or if it is dominated
by a dark-energy component, the above quoted requirements are
considerably relaxed. In particular, the standard dynamo can
produce the currently observed galactic magnetic fields from a
seed of the order of $10^{-30}$~G, or even less, at the end of
galaxy formation~\cite{DLT}. Note that a `collapsed' magnetic
field of $\sim10^{-30}$~G coherent on approximately $100$~pc
corresponds to a comoving field of the order of $10^{-34}$~G
spanning a scale of $\sim10$~kpc. Nevertheless, even seeds as week
as $10^{-34}$~G have been very difficult to produce in a
conventional way on the required scales. For example, assuming a
field with a coherence length of $10$~kpc and using
Eq.~(\ref{eq:r1}), we find a residual strength of approximately
$10^{-53}$~G. Clearly, such fields cannot seed the galactic dynamo
and are therefore astrophysically irrelevant.

The situation changes considerably if during inflation the
magnetic energy density decays as $a^{-2}$ instead of following
the adiabatic $a^{-4}$-law. As we have already seen, this happens
on the largest subcurvature scales (and beyond) when the
inflationary patch has negative spatial curvature. Therefore, the
universe can be permeated by substantially strong large-scale
magnetic fields even if it is only marginally open today. For a
direct comparison with the spatially flat case scenario, it helps
to follow the analysis of~\cite{TW} (see also Eq.~(\ref{eq:r1})
above). Consider a typical GUT-scale inflationary scenario with
$M\sim10^{17}$~GeV and reheating temperature $T_{\rm
RH}\sim10^9$~GeV. Then, for $B^2\propto a^{-2}$, the energy
density stored in a given magnetic mode at the end of inflation is
given by
\begin{equation}
\rho_{\rm B}\sim10^{-90}M^{8/3}T_{\rm RH}^{-2/3}
\tilde{\lambda}^{-2}_{\rm Mpc}\rho_{\gamma}
\sim10^{-51}\tilde{\lambda}^{-2}_{\rm Mpc}\rho_{\gamma}\,,
\label{eq:r2}
\end{equation}
instead of (\ref{eq:r1}). According to the above, on a given
scale, the earlier inflation starts and the lower the reheating
temperature, the stronger the supearadiabatically amplified
residual field. After inflation the high conductivity of the
plasma is restored. This ensures that $B^2\propto a^{-4}$ and
consequently that the ratio $r=\rho_{\rm B}/\rho_{\gamma}
\sim10^{-51}\tilde{\lambda}^{-2}_{\rm Mpc}$ remains fixed. To
proceed we note that $\tilde{\lambda}$ is nearly the curvature
scale at the end of inflation. Also, in a universe with nontrivial
spatial geometry the effect of curvature in a comoving region
remains unchanged, since the curvature scale simply redshifts with
the expansion (e.g.~see~\cite{LW}). This means that if
$1-\Omega_0$ is of the order of $10^{-2}$, as it appears to be
today~\cite{omega-today}, the current curvature scale is
\begin{equation}
\left(\lambda_k\right)_0=
\frac{\left(\lambda_H\right)_0}{\sqrt{1-\Omega_0}}\sim 10^4~{\rm
Mpc}\,,  \label{eq:lamdaH/k}
\end{equation}
where $\left(\lambda_H\right)_0=H_0^{-1}$,
$H_0\simeq2h\times10^{-42}$~GeV and $0.5\leq h\leq1$. The above is
also the approximate scale of the superadiabatically amplified
primordial magnetic field, redshifted to the present. Then, by
substituting this comoving scale into expression (\ref{eq:r2}) we
find that
\begin{equation}
r=\frac{\rho_{\rm B}}{\rho_{\gamma}}\sim10^{-59}\,,  \label{eq:B0}
\end{equation}
which corresponds to a magnetic field with current strength around
$10^{-35}$~G. Note that the above quoted strength depends on the
current values of the Hubble and the density parameters, although
this dependance is weak. Also, in order to satisfy the
conventional causality requirements we have implicitly assumed
that the universe was sufficiently open at the onset of inflation.
In particular, a relatively mild initial value of $\Omega_{\rm
i}<0.1$ will suffice for all practical purposes. Such a value
ensures that effectively all the largest subcurvature modes are
initially inside the horizon and therefore in causal contact when
inflation starts.

The first point to underline is that, to the best of our
knowledge, magnetic fields with $B_0\sim10^{-35}$~G and coherence
lengths of $\sim10^4$~Mpc are greatly stronger than any field
obtained within standard electromagnetic theory on such scales.
Moreover, fields with this strength are of astrophysical interest
because they can successfully seed the galactic dynamo, as long as
the current energy density of the universe is dominated by a dark
component; a scenario favoured by resent
observations~\cite{omega-today}. For a nearly flat universe with
the dark energy making up to 70\% of the present density
parameter, in particular, a seed field of $\sim10^{-35}$~G is
within the lower strength required for the galactic dynamo to
operate~\cite{DLT}. Note that the above given magnetic strengths
do not account for the effects of the physically more realistic
scenario of anisotropic protogalactic collapse. The latter is
expected to add a few more orders of magnitude to any field
obtained through the highly idealised spherical collapse
models~\cite{ZRS}.

For completeness, let us also consider the magnetic evolution on
supercurvature scales. During inflation supercurvature modes also
obey Eqs.~(\ref{eq:-1Bwave1}) and (\ref{eq:-1Bf}). On these scales
the eigenvalue $({\rm n})$ lies in the interval $[0,\,1)$, which
implies that $1<{\rm k}^2\leq2$. Then, near the ${\rm k}^2=2$
limit that corresponds to the homogeneous mode, the magnetic decay
rate becomes $B\propto a^{\sqrt{2}-2}$. The latter is considerably
slower than the $a^{-1}$-law associated with the largest
subcurvature scales. One should keep in mind, however, that
supercurvature scales in spatially open FRW cosmologies lie always
outside the Hubble radius and therefore are not causally
connected. Nevertheless, any magnetic field that happens to span
over these scales at the onset of inflation will decay much slower
than its subcurvature counterparts.

Finally, we should note that the linear amplification mechanism
outlined here, which is purely geometrical in nature, is in some
respects analogous to the one discussed in~\cite{M}. There, the
electromagnetic field is coupled to the inhomogeneous metric of a
perturbed FRW model. Given the a priori weakness of the field,
however, the magnetic amplification achieved in~\cite{M} is
presumably a nonlinear effect. The same can also be said about the
scenario discussed in~\cite{TDM}, where a weak primordial magnetic
field was amplified through its coupling to gravity wave
perturbations soon after the end of inflation.

\section{Discussion}
The origin and the evolution of the magnetic fields that we
observe almost everywhere in the universe today remains an open
issue and a matter of debate. The structure of the galactic
large-scale field strongly suggests a dynamo-type amplification
mechanism, but the latter requires a seed field to operate.
Depending on the efficiency of the large-scale dynamo, the
strength of the required seed varies between
$10^{-12}$-$10^{-23}$~G at the time of completed galaxy formation,
while its coherence length is approximately 10~kpc on comoving
scales. However, the questions regarding the origin of cosmic
magnetism involve not only the initial seed fields but the dynamo
mechanism itself. As yet, there is no final dynamo theory and the
whole subject is still under intense research~\cite{dynamo}.
Therefore, there is no certainty on what the properties of the
initial seed magnetic field should be. For instance, the fact that
astrophysical plasmas are gas mixtures (neutrals, ions and
electrons) can substantially modify the standard single fluid
approach (e.g.~see~\cite{MDBST}) and the dynamo
action~\cite{subramanian}. Besides, turbulent effects during the
radiation era can change the features of a primordial field by
enlarging, say, its coherent length~\cite{C}. Magnetic helicity is
also expected to play a pivotal role in these phenomena. Hence,
the requirements necessary for the subsequent MHD process that
will amplify the primordial seed could be substantially relaxed.

The geometry of our universe, whether it is open or closed, and
whether its energy density is close to the critical one is also an
open question of contemporary cosmology~\cite{CE}. Current
observations strongly suggest that the universe is nearly flat,
though they stop short from establishing whether it is marginally
open or marginally closed. It also appears that at present the
expansion dynamics is dictated by a dark-energy component, in the
form of a positive cosmological constant or quintessence. If so,
the standard constraints on the magnetic seed strength required
for the galactic dynamo to operate efficiently can be relaxed down
to $10^{-34}$~G, or even less. However, even fields as weak as
$10^{-34}$~G, on comoving scales of approximately $10$~kpc, are
very difficult to produce unless standard electromagnetism is
violated. The latter effectively means breaking the conformal
invariance of Maxwell's equations and in most of the cases this is
achieved by appealing to less well understood phenomenology. The
underlying reason is that in spatially flat FRW models the
magnetic fields decays as $B\propto a^{-2}$ always and
irrespective of plasma effects.

On these grounds, we have studied the evolution of cosmological
magnetic fields in perturbed FRW with nontrivial background
geometry. By allowing for curved spatial sections, we showed that
the adiabatic $B\propto a^{-2}$ law is not always guaranteed
because of the linear coupling between the field and the
background 3-geometry~\cite{T1}. When dealing with spatially open
FRW models, in particular, the extra curvature-related source term
in the magnetic wave equation meant that large-scale fields decay
as $a^{-1}$ instead of the standard adiabatic $a^{-2}$-law. This
is possible for fields evolving through a period of inflationary
expansion, due to the very low electrical conductivity of the
latter. As a result, primordial magnetic fields coherent on the
largest subcurvature scales could survive an epoch of inflation
and still be strong enough to sustain the dynamo process. Our
linear mechanism operates near the curvature scale and in
particular at the largest subcurvature scales. This in turn
ensures that the superadiabatically amplified magnetic field has
rather specific properties. Assuming that $1-\Omega\simeq10^{-2}$
today and that $H_0=100h$~km/sec$\cdot$Mpc, with $0.5\leq h\leq1$,
we find a residual field of the order of $10^{-35}$~G spanning
over a region of approximately $10^{4}$~Mpc. Magnetic fields like
these are by far stronger than any other large-scale field
obtained within standard electromagnetic theory. Moreover,
magnetic fields with the aforementioned properties are of
astrophysical interest provided the energy density of our universe
is currently dominated by a dark component. If so, a comoving
field of strength of the order of $10^{-35}$~G can seed the
large-scale galactic dynamo when its coherence scale is at least
as large as $10$~kpc. The latter is much less than the coherence
length of our superadiabatically amplified field, though we expect
fragmentation of the original seed field during the protogalactic
collapse and the subsequent nonlinear era.

If the universe is marginally open today, our mechanism allows for
a simple, viable and rather efficient amplification of large-scale
primordial seed magnetic fields to strengths that can seed the
galactic dynamo. Even if the universe is not open, this study
still brings about a rather important issue. This is the unique
nature and non-trivial properties of magnetic fields and their
potential implications in the context of general relativity.
Magnetic fields, in particular, are the only vector source that we
know that exist in the universe today and in the geometrical
framework of Einstein's theory vectors have different status than
scalars. The special status of the former, which is manifested in
the Ricci identities, couples the Maxwell field directly to the
geometry of the space in a natural way. This coupling has been
largely bypassed in the literature, though its implications are
generally non-trivial and in many cases quite
counter-intuitive~\cite{T2}. The best known example is probably
the scattering of electromagnetic waves by the gravitational
field, which leads to the violation of Huygens
principle~\cite{DWB}. Here, we have considered the implications of
this relativistic magneto-geometrical interaction for the
evolution of large-scale magnetic fields in FRW universes. We
found that, contrary to the widespread perception, a
superadiabatic-type amplification of cosmological magnetic fields
is possible in conventional cosmological models and within
standard electromagnetic theory. Therefore, in this case, the
magneto-geometrical coupling mimics effects that have been
traditionally attributed to new physics.

\section*{Acknowledgements}
We would like to thank Anthony Challinor, Anne Davis, Carlos
Martins, Kandu Subramanian, Shinji Tsujikava and Larry Widrow for
helpful discussions and comments.

\end{document}